\documentclass[runningheads]{llncs}
\usepackage{graphicx}
\usepackage{algorithm}
\usepackage[noend]{algpseudocode}
\usepackage{amsmath,amssymb,amsfonts}
\usepackage{multirow}
\usepackage{mathptmx} 
\usepackage{mathrsfs}
\usepackage{xcolor}
\usepackage[inline]{enumitem}

\begin{document}
\title{Detecting Online Hate Speech: Approaches Using Weak Supervision and Network Embedding Models \thanks{This paper is to appear in 2020 International Conference on Social Computing, Behavioral-Cultural Modeling \& Prediction and Behavior Representation in Modeling and Simulation (SBP-BRIMS'20)}}

\author{Michael Ridenhour\inst{1} \and
Arunkumar Bagavathi\inst{2} \and
Elaheh Raisi\inst{3} \and
Siddharth Krishnan\inst{1}}

\institute{UNC Charlotte
\email{\{mridenh7,skrishnan\}@uncc.edu} \and
Oklahoma State University
\email{abagava@okstate.edu} \and
Brown University
\email{elaheh\_raisi@brown.edu}}

\maketitle              
\begin{abstract}




The ubiquity of social media has transformed online interactions among individuals. Despite positive effects, it has also allowed anti-social elements to unite in alternative social media environments (eg. Gab.com) like never before. Detecting such hateful speech using automated techniques can allow social media platforms  to moderate their content and prevent nefarious activities like hate speech propagation. In this work, we propose a weak supervision deep learning model that - (i) quantitatively uncover hateful users and (ii) present a novel qualitative analysis to uncover \emph{indirect} hateful conversations.
This model scores content on the interaction level, rather than the post or user level, and allows for characterization of users who most frequently participate in hateful conversations. We evaluate our model on \emph{19.2M} posts and show that our weak supervision model outperforms the baseline models in identifying indirect hateful interactions. We also analyze a multilayer network, constructed from two types of user interactions in Gab(quote and reply) and interaction scores from the weak supervision model as edge weights, to predict hateful users. We utilize the multilayer network embedding methods to generate features for the prediction task and we show that considering user context from multiple networks help achieving better predictions of hateful users in Gab. We receive upto 7\% performance gain compared to single layer or homogeneous network embedding models.
\end{abstract}



\section{Introduction}
The widespread adoption of social media has transformed the way in which online interactions among individuals take place. Such interactions, say via tweets and re-tweets, often provide access to real-time news, viral marketing, online recruitment, etc. While the positive effects and applications of social media interactions are prominent, social media has also become a vehicle for degenerative behavior like cyber-bullying and hate speech propagation. Fringe outlets like Gab.com(shortly as Gab), under the guise of `free speech' have become channels for anti-social viewpoints like anti-Semitism and are able to galvanize supporters for the spread of hateful messages.
Therefore, it has become imperative to develop automated methods that can detect such messages and users who spread hateful ideologies to allow policymakers and stakeholders of social media sites to take appropriate measures to the spread of the anti-social content. 

For the purposes of this paper we focus on the topic of anti-Semitism in Gab. We choose this topic specifically because our dataset of choice for the modeling, Gab, has been shown to harbor users who are significantly more enthusiastic about posting anti-Semitic content than users on other sites \cite{kalmar2018twitter}. Furthermore, \cite{elsherief2018hate} has shown that religious-based hate speech tends to be targeted at an entire group, meaning we see more anti-Semitic content on the Jewish faith than at a specific person, thereby yielding a more comprehensive view on Gab users' vocabulary and beliefs surrounding the subject. 

There have been several advances in developing approaches to study hate speech propagation~\cite{zannettou2018gab}, but most approaches use direct markers that do not incorporate the context of the interaction. We define a weak supervision model, that takes a first step to address a specific gap in the research, by employing only very few lexicon labels to determine hate speech. We exploit a qualitative observation that the foundation of a lot of hate speech is based in content that does not overtly contain hate speech term, such as a slur. Our model is capable of detecting conversations containing nuances of hateful content, such as coded language, or hateful comments that do not contain previously identified hateful speech terms. We use the user interaction scores given by our model on user interaction networks to predict if a hateful or not. To this problem, we utilize multilayer network embedding frameworks~\cite{dong2017metapath2vec,bagavathi2018multi} to extract user features for the classification task.

In this paper we aim to answer following research questions to study hate speech and users in online social media:

\begin{itemize}
	\item \textbf{How to identify hate speech that are hidden in the meaning in social media interactions?} We define a weak supervision model and its corresponding function to qualitatively identify indirect hateful user interactions in Gab forum. 
	\item \textbf{How useful are user interaction networks to identify hateful users in the social media?} We define two weighted multilayer network random walk techniques: \emph{metapath} and \emph{multilayer} based methods to collect the node context. With features extracted using a skip-gram model, we define the user classification as a binary classification problem
	
	
\end{itemize}
\section{Related Work}
A significant research thrust has been formed to study the characteristics and dynamics of hate speech in Gab~\cite{zannettou2018gab} in past years. Hate speech encompasses a wide variety of topics and defining what exactly hate speech means can be an arduous task~\cite{zannettou2018gab}. Anti-Semitism and alt-right topics are specifically proven to have a growing influence in the Gab ecosystem~\cite{mcilroy2019welcome}. In an effort to understand how hate speech diffuses through Gab, a lexicon based approach~\cite{mathew2019spread} and belief network models~\cite{ribeiro2018characterizing} are available. Our model is adapted from a cyberbullying detection model that analyzes messages between users and outputs a representative score for the interaction \cite{fortuna2018survey,raisi2018weakly}. Another major challenge of defining hate speech is determining what language equates to hate speech since the use of a simple lexicon to identify hate speech can be imprecise and the vocabulary is ever changing~\cite{davidson2017automated,nobata2016abusive}. Our work address this issue by getting the sense of words based on their context. In the light of these limitations on lexicon-based approaches, we use feature representation of posts as a whole~\cite{le2014distributed,djuric2015hate}. 

Wide range of network embedding ideas have been introduced, including \emph{LINE}~\cite{tang2015line}, \emph{node2vec}~\cite{grover2016node2vec}, \emph{SDNE}~\cite{wang2016structural}, and \emph{GraphSAGE}~\cite{hamilton2017inductive}, after the success of skip-gram based word embedding model \textit{word2vec}~\cite{mikolov2013distributed}. Parallel to such research works on homogeneous networks, large number of research have been done on complex networks with multiple network layers, node types, and edge types~\cite{starnini2019communication}. Based on such complex connectivity patterns there exists network embedding models to extract features of nodes in the network, for example \emph{metapath2vec}~\cite{dong2017metapath2vec}, \emph{multi-net}~\cite{bagavathi2018multi}, and \emph{HIN}~\cite{shi2018heterogeneous}. In this work, we utilize such embedding algorithms on a multilayer network of user interactions with interaction scores, and perform node classification task on whether a user is hateful or not.


\section{Methodology}
\vspace{-1cm}
\begin{center}
\begin{figure*}
\label{fig:model_pipeline}
\centering
\includegraphics[scale=.46]{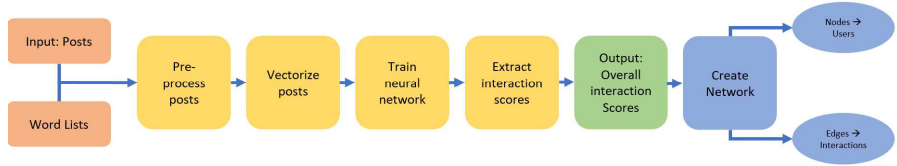}
\caption{Methodology overview - User interactions are quantified as interaction scores using the weak supervision model. These scores are given as edge weights to the multilayer network.}
\end{figure*}
\end{center}
\vspace{-1cm}

The technical aspect of our work can be two-fold: 1) Obtain user interaction scores using our weak supervision model which are capable to capture indirect markers of hate speech, and 2) Predict hateful users based on features extracted from multilayer networks built from multiple interaction types.

\subsection{Weak Supervision Model}



Weak supervision allows us to use far more data than we could label, providing more insights into such a large dataset.
The overview of our methodology is summarized in Figure~\ref{fig:model_pipeline}. We obtain interaction scores via an adaption of the \emph{Co-trained Ensemble model} introduced by~\cite{raisi2018weakly}. The original work co-train their model based on user interactions and network representations simultaneously, whereas we only use the textual representations from interactions. We employ such weak supervised model to use far more data than we could label, providing more insights into such a large dataset. The input of our model is a message/post $M$, and the output is a score indicating  the level of anti-Semitic hate incident in the message, i.e., $f: M \mapsto \mathbb{R}$. To train the model, we optimize the \emph{weakly supervised} loss function given in Equation~\ref{eq:weak_loss}.
	
\begin{equation}
\label{eq:weak_loss}
\begin{aligned}
\min_{\Theta} ~ \frac{1}{|M|} \sum_{m \in M} \ell \left( f(m; \Theta) \right)
\end{aligned}
\end{equation} 

where $\Theta$ is a model parameter, $l$ is the loss function, $f$ is the learning function, and $|M|$ is the total number of messages. Along with two set of keywords which are being used in the existing method: 1) anti-Semitic indicator words, 2) positive-sentiment words, we add \emph{context dependent words} also to the model. The anti-Semitic indicators are obvious hateful keywords. The context dependent words are associated with a topic, but are not considered hateful without a hateful context. Example words are ``jew'', ``rabbi'', and ``zionism''. Most of these terms are used in a negative context in Gab, but are not irrefutable hate speech terms on their own. 

Messages $m \in M$ is given a general range of bound values for what its ultimate score should be. The weak supervision is based on the fraction of values of aforementioned key-phrases. For a message $m \in M$ with $n$ total key phrases, if we assume $n^+(m)$ be the the  number  of  anti-Semitic  phrases, $n^-(m)$ be the number of positive-sentiment words, and $n^{\oplus}(m)$ is the number of context dependent words, then we bound the learning function by fraction of these indicators in the message $m$ as given in Equation~\ref{eq:bound}.

\begin{equation}
\label{eq:bound}
\begin{aligned}
\underbrace{2 \times \frac{n^+(m)}{n(m)} + \frac{n^{\oplus}(m)}{n(m)}}_{\textrm{Lower Bound}} < y_m < \underbrace{1 - \frac{n^-(m)}{n(m)}}_{\textrm{Upper Bound}} ,
\end{aligned}
\end{equation}

In the lower bound, we emphasize the anti-Semitic phrases with respect to the context dependent words by multiplying the fraction associated with anti-Semitic bound by 2. If the final score by the learner falls outside of the determined bounds, the model is penalized using weak supervision loss given in Equation~\ref{eq:weak_supervision_loss}.

\begin{equation}
\label{eq:weak_supervision_loss}
\begin{aligned}
\ell(y_m) &= -\log\left(\min\left\{1, 1 + (1 - \tfrac{n^-(m)}{n(m)}) - y_m\right\}\right) \\
&~~~~ - \log\left(\min\left\{1, 1 + y_m -\big(2 \times \tfrac{n^+(m)}{n(m)} + \tfrac{n^{\oplus}(m)}{n(m)}\big) \right\}\right)
\end{aligned}
\end{equation}
With these bounds, we use a linear neural network for our learner to transform the n-dimensional message representations, given by doc2vec, into a single score.

\subsection{Multilayer Network Embedding}
Multilayer networks can be formally defined as $G=\big(V_\textit{l},E_\textit{l}\big)_{\textit{l=1}}^{\mathscr{L}}$, where the network composes $\mathscr{L} > 1$ networks($G_1, G_2, ..., G_\mathscr{L}$) and each network($G_\textit{l}$) holds their own set of nodes($V_\textit{l}$) and edges($E_\textit{l}$). Although the definition of multilayer networks is similar to heterogeneous networks~\cite{dong2017metapath2vec}, multilayer networks have the property of single node types and multiple($\mathscr{L}$) edge types. Given that condition, each network in multilayer networks has some level of overlap in nodes (i.e) $\{ V_1 \cap V_2 \cap ... \cap V_\mathscr{L} \} \neq \O$. 



 
Given a multilayer network $G=\big(V_\textit{l},E_\textit{l}\big)_{\textit{l=1}}^{\mathscr{L}}$, where $\mathscr{L} > 1$, we utilize the skip-gram model based network embedding (\emph{node2vec})~\cite{grover2016node2vec} for the mutilayer context. Thus, we get n-dimensional vector representation for each node ($f_u$) based on their multilayered network structural patterns. We use the optimization function, given in Equation~\ref{eq:multilayer_learning}, similar to the ones proposed in \textit{Metapath2Vec}~\cite{dong2017metapath2vec} and \textit{Multi-Net}~\cite{bagavathi2018multi} frameworks to maximize observing the neighborhood $N(u)$ across all network layers conditioned on the node $u$. We give in-depth analysis on how these vector space features are utilized for hateful user classification in Section~\ref{sect:exp_classi}

\begin{equation}
	\label{eq:multilayer_learning}
	\begin{aligned}
		& \underset{\textit{f}}{\text{\textit{max}}}
		& & \sum_{u \in V}^{} \sum_{\textit{l} \in \mathscr{L}} \sum_{v \in \textit{N}(u)_\textit{l}}^{} log \  Pr \big(v_\textit{l} | f_u \big)
	\end{aligned}
\end{equation}

	, where $\textit{N}(u)_\textit{l}$ is the neighborhood of node $u$ at the $\textit{l}^{th}$ network and we model the $Pr \big(v_\textit{l} | f_u \big)$ using a softmax function.
	
\subsubsection{Random Walks}
\textit{Random walks} are used to define a node's neighborhood or context for node embedding models. Although, several types of random walks are available, we explore random walks defined specifically for multilayer networks~\cite{dong2017metapath2vec,bagavathi2018multi} and give the context of edge weights for each walk. In this work, we employ two strategies: 1) metapath based random walk and 2) multilayer random walk.




\subsubsection{Metapath based random walks}
Metapath based random walks are used in heterogeneous networks~\cite{dong2017metapath2vec}, where there exists multiple types of nodes and edges. In metapath based random walks, a structural schema($\mathcal{S}$) is designed to bias the random walker to visit all the network layers. We define a metapath schema as: $\mathcal{S}=E_1 \rightarrow E_2 \rightarrow ... \rightarrow E_\textit{l} \rightarrow E_1$, where $E_i$ is an edge type. Given a multilayer network $G$ and a metapath schema $\mathcal{S}$, we define the transition probability for each edge as given in Equation~\ref{eq:metapath_transition}

\begin{equation}
	\label{eq:metapath_transition}
	\begin{aligned}
	P \big( V_{i+1} \ | \ V_{i} \big) \ = \ \left\{
	                \begin{array}{ll}
	                  w_{V_{i}}^{V_{i+1}} \ if \ E \big(V_i,V_{i+1}\big) \in E_{t+1} \\
	                  0 \  \ \ \ \ \ otherwise
	                \end{array}
	              \right.
	\end{aligned}
\end{equation}

	where $w_{V_{i}}^{V_{i+1}}$ is the normalized edge weight and $t \in \mathcal{S}$ denotes the current state in the metapath schema.

\subsubsection{Multilayer random walks}
In multilayer random walks, we relax the idea of metapath schema and the random walker is allowed to choose 2 options at a time step $i$: 1) stay in the same network with probability $\textit{p}$ and visit the next node $V_{i+1}$ or 2) shift to a random network layer with probability $1 - \textit{p}$ and visit $V_{i+1}$. The transition probability for visiting the next node is given in Equation~\ref{eq:multinet_transition}. A transition to another network layer(say $G_n$ from $G_m$) happens if and only if the node $V_i$ is available in $G_n$.

\begin{equation}
	\label{eq:multinet_transition}
	\begin{aligned}
	P \big( V_{i+1} \ | \ V_{i} \big) \ = \ \left\{
	                \begin{array}{ll}
	                  w_{V_{i}}^{V_{i+1}} \ if \ E \big(V_i,V_{i+1}\big) \in E_{\textit{l=1}}^{\mathscr{L}} \\
	                  0 \  \ \ \ \ \ otherwise
	                \end{array}
	              \right.
	\end{aligned}
\end{equation}
\section{Experiments and Results}
\subsection{Dataset Description}
We utilize a dataset sample collected from \emph{Gab.com} containing 19,127,608 posts \cite{fair2019shouting} for our experiments and observations. The distribution of posts is given in Table~\ref{tab:ds_props} and a representative sample of our indicator lexicons is given in Table~\ref{tab:lexicons}. The highest concentration of anti-Semitic indicator terms is in quotes(similar to re-tweet in Twitter), with 0.36\% more instances than replies and almost three times more instances than original posts. Interestingly, these types of posts are originally sourced from other users. This coupled with the higher concentration of hateful terms in replies points to the observation that Gab users are more hateful or exposed to more hateful content when interacting with other users than they are when posting original content to their followers. This observation, coupled with the findings that posts from hateful users spread more than posts from non-hateful users \cite{mathew2019spread}, lead us to believe that detecting hateful content at the interaction level will provide valuable information when classifying hateful users.

\begin{table}[h!]
	\begin{center}
		\caption{Properties of the Gab Dataset - Numbers indicate the quantity of each post type in the total dataset. Percentages indicates the volume of anti-Semitic indicators in each type of post.}
		\label{tab:ds_props}
		\begin{tabular}{|c|c|c|}
			\hline
			\textbf{Post Type} & \textbf{\# in Dataset}  & \textbf{\% with direct indicator(s)} \\
			\hline
			Quote & 2,321,744 & 1.11\% \\
			Reply & 6,916,904 & 0.75\% \\
			Original & 9,888,960 & 0.38\% \\
			\hline
		\end{tabular}
	\end{center}
\end{table}
\begin{table}[ht]
\caption{Example lexicons used to determine the bounds for each post(see Methodology)}
\label{tab:lexicons}
\begin{center}
\begin{tabular}{|c|c|}
    \hline
    \multirow{3}{*}{\textbf{Anti-Semitic Indicators}}&'k*ke'\\
    &'\#hitlerwasright'\\
    &'h*eb'\\
    \hline
    \multirow{3}{*}{\textbf{Context Dependent Phrases}}&'jew'\\
    &'talmud'\\
    &'zionism'\\
    \hline
\end{tabular}
\end{center}
\end{table}

\begin{figure*}
\label{fig:score_dist}
\centering
\includegraphics[scale=.35,width=5.5cm]{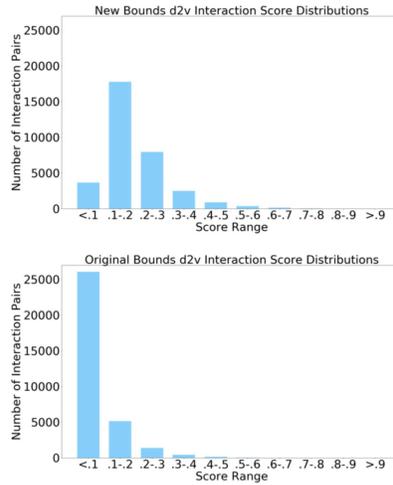}
\caption{Distributions of Scores for each variant of the model and embedding options.}
\end{figure*}

\begin{table}[ht]
\caption{Example direct and indirect anti-Semitism interactions. Direct interaction contains a hateful key word, while indirect interaction contains hateful content as an insulting language.}
\label{tab:example_interactions}
\begin{center}
\begin{tabular}{|p{5cm}|p{5cm}|}
	\hline
	\textbf{Direct hate speech example} & \textbf{Indirect hate speech example} \\
	\hline
	\textcolor{blue}{User 1:} \indent \textit{because mo****kers like you think israel should die off!!!} \newline
	\textcolor{red}{User 2:} \indent \textit{if jews been kicked out ... f**ing pathetic with your keyboard temper tantrums there, jew b**ch} \newline
	\textcolor{red}{User 2:} \indent \textit{and you do the k*kes work so well !} \newline
	\textcolor{red}{User 2:} \indent \textit{because mother f**kers like me are sick of usury , religious bigots,...} &
	
	\textcolor{red}{User 2:} \indent \textit{i say we expel ((them)) all to israel, and then flood israel with muslims. let them enjoy some forced "diversity"} \newline
	\textcolor{blue}{User 1:} \indent \textit{....they have to be probed for jewish connections. this is a ritual psyop preying on white children.} \newline	
	\textcolor{blue}{User 1:} \indent \textit{.......they control america with \#pedo \#f*ggots.} \\
	\hline
\end{tabular}
\end{center}
\end{table}

\subsection{Weak Supervision Model}

We examine interactions and interaction scores from the weak supervision model~\cite{raisi2018weakly} to get useful information from user posts. We first generate doc2vec representations of the posts using both our modified bounds function given in Equation~\ref{eq:bound}, and we use the original bounds function which excludes the presence of the context dependent term list and weight the indicator list and positive words list equally as a baseline model. Figure~\ref{fig:score_dist} gives the distribution of scores for each model variant, indicating that the majority of posts have little anti-Semitic content. To ensure the model will suffice our multilayer network embedding model, we qualitatively look at the distribution of scores and the textual content of the top 100 most hateful interactions produced by each bounds and representation variant.

By studying 100 interactions we notice two types of anti-Semitic hate speech: \emph{direct} and \emph{indirect}. Direct hate speech contains a word or hashtag from our lexicon of anti-Semitic indicators, while indirect hate speech does not contain any of our anti-Semitic indicators, but often link a word from our context dependent word list to a hateful content. Example direct and indirect hate interactions are given in Table~\ref{tab:example_interactions}. Both of these interactions are sourced from the top 100 posts using the new bounds method and doc2vec representations. The ability of this model to pick up on examples of indirect hate speech further shows this model is useful when detecting hateful users. 

\subsection{User classification}
\label{sect:exp_classi}
We predict hateful users from their structural patterns in reply and quote (re-tweet in Twitter) networks of Gab. We represent these networks as a weighted 2-layer multilayer network, where network vertices are user accounts, network edges are user replied to/quoted another user, and edge weight represent interaction score from the weak supervision model. Summary of our networks is given in Table~\ref{tab:network_statistics}. As the Gab data users are unlabeled, we replicate the methods used by \cite{mathew2019spread} to label a user hateful if they have used words in our anti-Semitic indicators list at a frequency of $z$. In our experiments, we set $z=5$.

\begin{table}[!t]
\caption{Network statistics}
\label{tab:network_statistics}
\centering

\begin{tabular}{|c|c|c|}
\hline
\hline
\textbf{Property} & \textbf{Reply network} & \textbf{Quote network} \\
\hline
\hline
\# nodes & 9,813 & 4,069 \\
\hline
\# edges & 45,728 & 16,836 \\
\hline
Avg. degree & 4.66 & 4.14 \\
\hline
\# components & 174 & 35 \\
\hline
\end{tabular}

\end{table}

\begin{table*}
\caption{Binary classifier results - Average of 10 F1 measures. \textcolor{green}{Green} - best model performance and \textcolor{blue}{Blue} - second best model performance.}
\label{tab:embedding_results}

\centering
\begin{tabular}{|c|c|c|c|c|c|c|c|c|c|c|c|}
\hline
\textbf{Method type} & \textbf{Algorithm} & \textbf{Merge type} & \multicolumn{9}{c}{\textbf{Training ratio}} \\
\hline
& & & 10\% & 20\% & 30\% & 40\% & 50\% & 60\% & 70\% & 80\% & 90\% \\
\hline
\multirow{2}{*}{Flatten networks} & node2vec & - & 0.48 & 0.47 & 0.46 & 0.46 & 0.46 & 0.46 & 0.46 & 0.56 & 0.49 \\\cline{2-12}
 & LINE & - & 0.61 & 0.61 & 0.63 & 0.65 & 0.66 & 0.66 & 0.65 & 0.65 & 0.66\\
\hline
\hline
\multirow{8}{*}{Merge embeddings} & \multirow{4}{*}{node2vec} & Average & 0.67 & 0.67 & 0.67 & 0.68 & 0.68 & 0.71 & 0.71 & 0.68 & 0.68 \\\cline{3-12}
 & & Max-pooling & 0.65 & 0.65 & 0.65 & 0.66 & 0.67 & 0.71 & 0.70 & 0.68 & 0.69 \\\cline{3-12}
 & & Gated Reply & 0.42 & 0.43 & 0.43 & 0.43 & 0.43 & 0.45 & 0.45 & 0.53 & 0.47 \\\cline{3-12}
 & & Gated Quote & 0.66 & 0.67 & 0.69 & 0.70 & 0.70 & 0.71 & 0.72 & 0.70 & 0.70 \\\cline{2-12}
 & \multirow{4}{*}{LINE} & Average & 0.66 & 0.64 & 0.62 & 0.62 & 0.62 & 0.63 & 0.64 & 0.64 & 0.64 \\\cline{3-12}
 & & Max-pooling & 0.60 & 0.60 & 0.61 & 0.61 & 0.63 & 0.65 & 0.65 & 0.65 & 0.63 \\\cline{3-12}
 & & Gated Reply & 0.33 & 0.32 & 0.30 & 0.41 & 0.47 & 0.50 & 0.50 & 0.50 & 0.45 \\\cline{3-12}
 & & Gated Quote & 0.63 & 0.64 & 0.64 & 0.66 & 0.66 & 0.67 & 0.68 & 0.64 & 0.63 \\\cline{2-12}
\hline
\hline
\multirow{2}{*}{Multilayer embedding} & Metapath2vec & - & \textcolor{blue}{0.70} & \textcolor{blue}{0.73} & \textcolor{blue}{0.72} & \textcolor{blue}{0.73} & \textcolor{blue}{0.75} & \textcolor{blue}{\textbf{0.77}} & \textcolor{blue}{\textbf{0.77}} & \textcolor{blue}{0.75} & \textcolor{blue}{0.75} \\\cline{2-12}
 & MultiNet & - & \textcolor{green}{0.72} & \textcolor{green}{0.74} & \textcolor{green}{0.74} & \textcolor{green}{0.75} & \textcolor{green}{0.77} & \textcolor{green}{\textbf{0.79}} & \textcolor{green}{0.78} & \textcolor{green}{0.76} & \textcolor{green}{0.76} \\
\hline
\end{tabular}
\end{table*}


We define the user classification problem as a binary mapping function: $\textit{f}: \ f_u \mapsto 0|1$, where $f_u$ is a feature vector of user $u$ obtained from network embedding and labels $0$ and $1$ defines that the user is not-hateful and hateful respectively. 

\subsubsection{Baseline setup}
We utilize 2 random-walk based frameworks: \textit{node2vec}~\cite{grover2016node2vec} and \textit{Line}~\cite{tang2015line} as baseline models, which support both directed and weighted graphs.

\begin{itemize}
	\item \textbf{\textit{node2vec}} (\emph{n2v}): This method uses biased random walk to explore the neighborhood. We set the parameters $p=1$ and $q=1$ for their random walks
	\item \textbf{\textit{LINE}}: This method aims to maximize the second order proximity of nodes to collect their neighborhood.
\end{itemize}


Since the above-mentioned methods are designed for single layer networks, we strategize in following methods to combine feature representations from the multilayer setup.


\begin{itemize}
	\item \textbf{Flatten networks}: We collapse all network layers into one network. If there exists multiple edges, we take average of their edge weights and normalize the weights.
	\item \textbf{Merge embeddings}: Embedding from two models are merged using the function $f_v = w_1\big(f_{v_a}\big) \: o \: w_2\big(f_{v_b}\big)$, where $f_{v_a}$ and $f_{v_b}$ are feature vectors of the node $v$ from layers $a$ and $b$ respectively, $o$ is one of the component-wise merge operation called \emph{Gated embedding}, where we apply a simple sigmoid function($\sigma$) to combine vectors, and $w_1$ and $w_2$ are weighted operations over corresponding vectors~\cite{kiela2018efficient}.
\end{itemize}

We use both \textit{node2vec} and \textit{LINE} along with all the above mentioned embedding generation methods. User embeddings are passed as inputs to the binary classifier. We give the \textit{F1 measure} of the model for varying training data size in Table~\ref{tab:embedding_results}. Overall, we get significant performance gain (atleast \textit{7\%}) with the multilayer based network embedding methods(\emph{metapath2vec} and \emph{multi-net}) compared to other two methods(flatten networks and merge embeddings). Out of the two models, we show atleast \textit{2\%} performance improvement with the multi-net based random walks. We also can evidence that flattening the networks does not yield good results with both \emph{node2vec} and \emph{LINE} models, while they give varied performance with different merge strategies. Even though we have given results based on weighted graphs only, we found atleast 10\% performance decrease with embedding using unweighted networks and are not reported due to space limitations. The performance gain with weighted networks is due to collecting better node context based on interaction scores. Figure 3 give models performance in terms of accuracy, precision, and recall. For these results, we fix the training sample size to be \textit{60\%} and average all results from 10 iterations. 




\begin{figure}
	\centering
	\label{fig:model_performance}
	\includegraphics[width=0.5\textwidth]{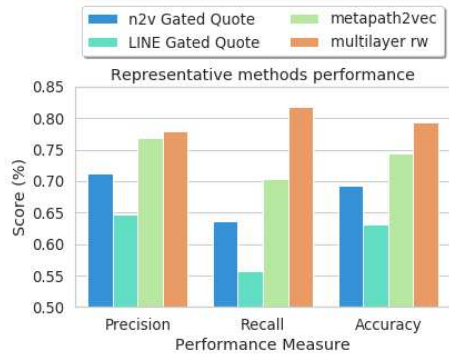}
	\caption{Performance measure of basline and proposed methods. Each measure is averaged over 10 times.}
\end{figure}
\vspace{-1cm}
\section{Conclusion and Discussion}
As the language of anti-Semitism, racism, white supremacy, and other forms of hate speech evolves and spreads through social media, it is of the utmost importance that we develop tools to monitor its proponents. Using methodologies defined as weak supervision and multilayer network embedding model, we have provided a qualitative way to identify indirect hate speech and quantitatively classify hateful users in Gab based on their neighborhood patterns. With our experiments, we showed that multilayer based classification performs atleast $7\%$ better than baseline models. This work involve multiple scopes in the future including: \begin{enumerate*}
	\item Quantitative evaluation of indirect hate speech detection in user conversations
	\item Systematic update and maintenance of more precise lexicons to increase the performance of this model,
	\item Add a transfer learning aspect, where we model interactions propagate between social media forums,
	\item Design a multi-modal classification algorithm to use multiple user features like their network properties, language, information they consume and share to perform the user classification task
\end{enumerate*}

%
%
\bibliographystyle{splncs04}
{\footnotesize
\bibliography{bibliography}}

\begin{thebibliography}{10}
\providecommand{\url}[1]{\texttt{#1}}
\providecommand{\urlprefix}{URL }
\providecommand{\doi}[1]{https://doi.org/#1}

\bibitem{bagavathi2018multi}
Bagavathi, A., Krishnan, S.: Multi-net: A scalable multiplex network embedding
  framework. In: International Conference on Complex Networks and their
  Applications. pp. 119--131 (2018)

\bibitem{davidson2017automated}
Davidson, T., Warmsley, D., Macy, M., Weber, I.: Automated hate speech
  detection and the problem of offensive language. In: AAAI ICWSM (2017)

\bibitem{djuric2015hate}
Djuric, N., Zhou, J., Morris, R., Grbovic, M., Radosavljevic, V., Bhamidipati,
  N.: Hate speech detection with comment embeddings. In: ACM WWW. pp. 29--30
  (2015)

\bibitem{dong2017metapath2vec}
Dong, Y., Chawla, N.V., Swami, A.: metapath2vec: Scalable representation
  learning for heterogeneous networks. In: ACM SIGKDD. pp. 135--144 (2017)

\bibitem{elsherief2018hate}
ElSherief, M., Kulkarni, V., Nguyen, D., Wang, W.Y., Belding, E.: Hate lingo: A
  target-based linguistic analysis of hate speech in social media. In: AAAI
  ICWSM (2018)

\bibitem{fair2019shouting}
Fair, G., Wesslen, R.: Shouting into the void: A database of the alternative
  social media platform gab. In: AAAI ICWSM. pp. 608--610 (2019)

\bibitem{fortuna2018survey}
Fortuna, P., Nunes, S.: A survey on automatic detection of hate speech in text.
  ACM Computing Surveys (CSUR)  \textbf{51}(4), ~85 (2018)

\bibitem{grover2016node2vec}
Grover, A., Leskovec, J.: node2vec: Scalable feature learning for networks. In:
  ACM SIGKDD. pp. 855--864 (2016)

\bibitem{hamilton2017inductive}
Hamilton, W., Ying, Z., Leskovec, J.: Inductive representation learning on
  large graphs. In: NIPS. pp. 1024--1034 (2017)

\bibitem{kalmar2018twitter}
Kalmar, I., Stevens, C., Worby, N.: Twitter, gab, and racism: the case of the
  soros myth. In: ACM International Conference on Social Media and Society. pp.
  330--334 (2018)

\bibitem{kiela2018efficient}
Kiela, D., Grave, E., Joulin, A., Mikolov, T.: Efficient large-scale
  multi-modal classification. In: Thirty-Second AAAI Conference on Artificial
  Intelligence (2018)

\bibitem{le2014distributed}
Le, Q., Mikolov, T.: Distributed representations of sentences and documents.
  In: ICML. pp. 1188--1196 (2014)

\bibitem{mathew2019spread}
Mathew, B., Dutt, R., Goyal, P., Mukherjee, A.: Spread of hate speech in online
  social media. In: ACM Web Science. pp. 173--182 (2019)

\bibitem{mcilroy2019welcome}
McIlroy-Young, R., Anderson, A.: From “welcome new gabbers” to the
  pittsburgh synagogue shooting: The evolution of gab. In: AAAI ICWSM. pp.
  651--654 (2019)

\bibitem{mikolov2013distributed}
Mikolov, T., Sutskever, I., Chen, K., Corrado, G.S., Dean, J.: Distributed
  representations of words and phrases and their compositionality. In: NIPS.
  pp. 3111--3119 (2013)

\bibitem{nobata2016abusive}
Nobata, C., Tetreault, J., Thomas, A., Mehdad, Y., Chang, Y.: Abusive language
  detection in online user content. In: ACM WWW. pp. 145--153 (2016)

\bibitem{raisi2018weakly}
Raisi, E., Huang, B.: Weakly supervised cyberbullying detection using
  co-trained ensembles of embedding models. In: IEEE/ACM ASONAM. pp. 479--486
  (2018)

\bibitem{ribeiro2018characterizing}
Ribeiro, M.H., Calais, P.H., Santos, Y.A., Almeida, V.A., Meira~Jr, W.:
  Characterizing and detecting hateful users on twitter. In: AAAI ICWSM (2018)

\bibitem{shi2018heterogeneous}
Shi, C., Hu, B., Zhao, W.X., Philip, S.Y.: Heterogeneous information network
  embedding for recommendation. IEEE Transactions on Knowledge and Data
  Engineering  \textbf{31}(2),  357--370 (2018)

\bibitem{starnini2019communication}
Starnini, M., Boguñá, M., Serrano, M.n.: The interconnected wealth of
  nations: Shock propagation on global trade-investment multiplex networks.
  Scientific Reports  \textbf{9}(1) (2019)

\bibitem{tang2015line}
Tang, J., Qu, M., Wang, M., Zhang, M., Yan, J., Mei, Q.: Line: Large-scale
  information network embedding. In: ACM WWW. pp. 1067--1077 (2015)

\bibitem{wang2016structural}
Wang, D., Cui, P., Zhu, W.: Structural deep network embedding. In: ACM SIGKDD.
  pp. 1225--1234 (2016)

\bibitem{zannettou2018gab}
Zannettou, S., Bradlyn, B., De~Cristofaro, E., Kwak, H., Sirivianos, M.,
  Stringini, G., Blackburn, J.: What is gab: A bastion of free speech or an
  alt-right echo chamber. In: ACM WWW. pp. 1007--1014 (2018)

\end{thebibliography}
\end{document}